# THE EFFECTS OF TIME SINCE FIRE ON BIRD COMMUNITY COMPOSITION IN CHAPARRAL ECOSYSTEMS ACROSS LOS ANGELES COUNTY

Lucas Qiu, Daniel Stockel, James Kraynik, Katie Lau, Ashley Yoon
University of California, Los Angeles
Department of Ecology & Evolutionary Biology

## Abstract

This study investigates the impact of time since fire on bird community composition in Southern California chaparral ecosystems. We surveyed avian richness and abundance across 14 sites representing a 0–25-year post-fire chronosequence in Los Angeles County. Sites burned within the last five years supported fewer species, primarily dominated by generalists, while mid- to late-successional sites exhibited greater richness and a higher proportion of specialists. These patterns corresponded with increases in vegetation structural complexity over time. However, no consistent relationships were found between bird communities and abiotic variables, such as weather, temperature, and elevation, likely due to the single-visit sampling design. Our results align with successional theory and underscore the ecological importance of fire return intervals that allow full chaparral recovery. Restoration and management should prioritize long-term structural development, invasive grass control, and post-fire heterogeneity to support diverse and resilient avian communities.

# 1. Introduction

Understanding how disturbances, such as wildfires, alter community composition, particularly in biodiversity-rich and fire-adapted ecosystems like Southern California chaparral, is crucial for predicting long-term ecological resilience and survival. This study investigates the impact of post-fire changes in vegetation structure on bird species richness and community composition across a chronosequence of recovery. Specifically, we investigate the extent to which vegetation complexity supports specialist versus generalist bird species and how these dynamics shift across a time-since-fire gradient. These insights are particularly relevant given the ongoing changes in fire regimes and habitat transformation in Mediterranean climate systems. In this context, the intermediate disturbance hypothesis (Connell 1978) offers a useful framework, suggesting that species diversity is highest at intermediate levels of disturbance, where both early colonizers and later successional species can coexist. Avian communities serve as valuable indicators of these shifts, as their richness and composition closely track vegetation complexity and habitat quality.

To contextualize this research, we explore the broader ecological principle of terrestrial disturbance. These disturbance events, whether natural, such as wildfires and floods, or anthropogenic, including urbanization and land-use change, reshape ecosystems by altering microclimate, vegetation composition, and resource availability (Pickett and White 1985, Turner 2010, Schoennagel et al. 2017). The aftermath of disturbance often triggers complex ecological cascades that necessitate either adaptation or exclusion for affected species. Among these forces, fire stands out as one of the most ecologically significant. In terrestrial systems, it reshapes nutrient cycling, habitat availability, and species interactions (Bond and Keeley 2005). In fire-adapted landscapes like Southern California chaparral, these periodic disturbances have historically supported ecological resilience and promoted biodiversity through successional turnover. Consequently, chaparral ecosystems provide a valuable model for studying post-disturbance dynamics.

Chaparral habitats (comprising Coastal Sage and Chaparral, Montane Chaparral and Woodlands, and Interior Chaparral and Woodlands) are defined by high plant endemism and adaptation to periodic wildfire, with natural return intervals historically ranging from 30 to 100 years (Barro and Conard 1991, Keeley et al. 2008). These systems are dominated by species with fire-resilient traits like basal resprouting, smoke-stimulated germination, and serotiny (Keeley and Pausas 2022), which enable them to recover naturally under historical fire regimes.

However, recent decades have brought increasingly frequent and severe fires, driven by climate warming, land-use changes, fire suppression policies, and the expansion of the wildland–urban interface, a major issue in the recent Los Angeles Wildfires (Syphard et al. 2009, Taccaliti et al. 2023, Kimball et al. 2024, Qiu et al. 2025). These disruptions have accelerated a process known as type conversion, wherein invasive grasses outcompete native shrubs and alter successional trajectories (Pratt 2022). As a result, habitat complexity declines, threatening ecosystem functions and species that depend on native vegetation structure (Gaertner et al. 2014, Glassman et al. 2023).

This degradation has serious consequences for fauna reliant on late-successional habitats. Many obligate and facultative species, including shrub-nesting birds, depend on dense woody vegetation for nesting, cover, and foraging (Barbour 2007). Disruption of the fire cycle not only fragments these habitats but also compresses successional stages, disproportionately affecting species with specialized requirements (Keeley et al. 2008). Recovery from fire is not spatially uniform, and heterogeneity in topography and soil conditions further complicates regeneration patterns. Slope, aspect, and microclimate influence the pace and trajectory of vegetation recovery, shaping which plant and animal species recolonize a site and when (Van Mantgem et al. 2015). These spatial dynamics emphasize the need for research designs that compare multiple post-fire sites along a temporal gradient.

Birds, with their rapid response to habitat changes and strong ties to vegetation structure, are well-suited to assess post-disturbance recovery. In Mediterranean systems, avian communities typically follow predictable successional patterns: generalists dominate shortly after fire, while species richness and specialization increase as vegetation matures and structural complexity improves (Mendelsohn et al. 2008, Clavero et al. 2011, Hargrove and Unitt 2018). Structural attributes like vertical stratification, which create microhabitats across vegetation layers, are particularly important for supporting diverse bird communities (MacArthur and MacArthur 1961, Moreira et al. 2003).

Yet species richness alone cannot fully capture the quality of ecological recovery. Richness may increase even as abundance declines or community composition shifts toward generalists (Julliard et al. 2006, Clavero et al. 2011). Functional diversity, foraging guilds, and habitat specialization offer more nuanced insights into community integrity (Gregory and van Strien 2010, Saavedra et al. 2015). These patterns are further complicated by biotic and abiotic filters, such as invasive grasses, which inhibit the regrowth of native shrubs and suppress the return of structurally complex vegetation needed by many bird species (Park and Jenerette 2019a, Pratt 2022, Pulido-Chavez et al. 2023). Moreover, the timeline for bird community recovery is often protracted. Some species show delayed recolonization due to dispersal limitations, structural habitat requirements, or interspecific interactions. These asynchronous patterns between vegetation recovery and bird response can mislead short-term assessments, underscoring the value of long-term ecological monitoring (Brawn et al. 2001, Kelly et al. 2020).

Despite these challenges, Southern California chaparral remains an essential system for studying disturbance ecology. The tight coupling between vegetation and fauna, coupled with increasing fire pressures, presents both an urgent conservation challenge and a scientific opportunity. Insights from this system have broad relevance for fire management, habitat restoration, and biodiversity conservation in fire-prone Mediterranean-type regions globally.

To address these issues, our study utilizes spatial variability in time since fire across 14 chaparral-dominated sites in Los Angeles County. By applying a space-for-time substitution approach, we examine how vegetation structure mediates bird species richness and community composition along a 25-year post-fire chronosequence. We hypothesized that recently burned sites, characterized by low vegetation complexity, would exhibit reduced richness, with diversity increasing over time in concert with canopy height, shrub density, and vertical layering. Through paired avian point counts and vegetation surveys, we aim to clarify how ecological succession shapes bird communities and inform strategies for post-fire habitat recovery.

## 2. Methods

*2.1 Site Selection and Preparation*

Fieldwork was conducted between April 24 and May 16, 2025, across 14 chaparral-dominated sites in Los Angeles County, California. These sites spanned Coastal Sage and Chaparral, Montane Chaparral and Woodlands, and Interior Chaparral and Woodlands ecoregions. Using data layers overlaid in qGIS, sites were selected based on recent wildfire history sourced from the CalFire Fire and Resource Assessment Program (FRAP), and the Los Angeles County Geohub. **Table 1** outlines twelve sites that had burned within the past 25 years and met the following criteria: fire size ≥ 100 acres, dominance by chaparral vegetation, proximity to roads (<500 m), and ≥ 1 km distance from WUI.

A YEAR_MAX filter was applied to exclude sites with subsequent overlapping burns. Two additional sites were selected as controls, having remained unburned for at least 30 years, based on the historical fire-return interval for chaparral (Van de Water & Safford, 2011). **Figure 1** places the locations across Los Angeles County based on the initial transect coordinates for each site.

| Access Location | Fire Name | Fire Year | Coordinates |
|---|---|---|---|
| Devils Canyon Trail | Simi Fire | 2003 | 34.15855, -118.36338 |
| Wilson Canyon Trail (MRCA Open Space) | Sayre Fire | 2008 | 34.1955, -118.2623 |
| George's Gap Trail (Angeles Crest NF) | Station Fire | 2009 | 34.16232, -118.09920 |

| Van Tassel Trail | Fish Fire | 2016 | 34.09396, -117.56255 |
| --- | --- | --- | --- |
| Anza Loop Trail/Old 101 Trail (Mustard) | Woolsey Fire | 2018 | 34.14481, -118.68976 |
| Crags Road (Malibu Creek SP) | Woolsey Fire | 2018 | 34.559, -118.4357 |
| Azusa Cyn West Fork Trail (San Gabriel River) | Bobcat Fire | 2020 | 34.1433, -117.5216 |
| Santa Ynez Canyon Trail | Palisades Fire | 2021 | 34.05425, -118.34737 |
| Fisherman Trail (Castaic) | Route Fire | 2022 | 34.31425, -118. 37106 |
| Westridge Canyonback Trail | Palisades Fire | 2025 | 34.541, -118.3047 |
| Castaic Lake Campground | Hughes Fire | 2025 | 34.3029, -118. 3622 |
| Altadena Crest Trail | Eaton Fire | 2025 | 34.1228, -118.853 |
| King Gillette Ranch | N/A | Control | 34.612, -118.4211 |
| Hollywood Sign - Mulholland Trail | N/A | Control | 34.754, -118.1856 |

**Table 1.** *Summary of study sites surveyed across fire recovery gradients.* Includes site names, associated fire event, year of most recent burn, and GPS coordinates of initial transect point in Decimal Degree format (Latitude, Longitude). Fire years span from 2003-2025 (not including control sites), representing four successional stages plus unburned controls.

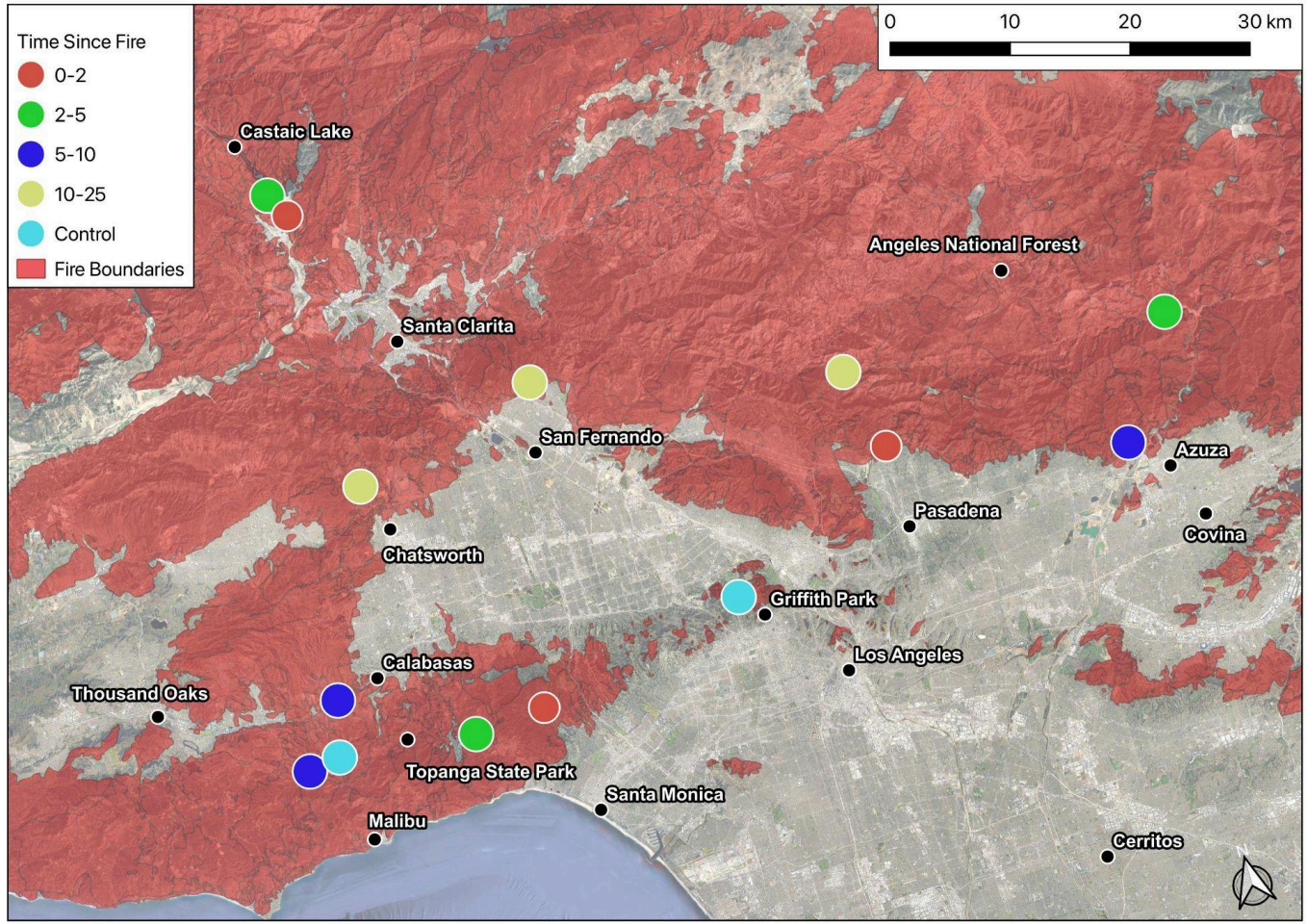

**Figure 1.** *Map of selected study sites across Los Angeles County.* Locations represent a chronosequence of 14 chaparral-dominated sites ranging from 0-25 years post-fire, including two 30+ year unburned control sites. Sites were selected using GIS analysis of fire history, vegetation, and accessibility constraints.

## *2.2. Sampling Design*

Survey points were generated using qGIS's "Random Selection Within Subsets" and "Points Along Geometry" tools. Trails intersecting fire perimeters were clipped to boundary extent, and random points were generated within 500 m of trails and roads. Sites were stratified into four time-since-fire categories: 0–2 years, 2–5 years, 5–10 years, and 10–25 years. Up to six candidate points were selected per category and from these, three were randomly chosen for field sampling.

## *2.3. Field Data Collection*

At each site, three survey locations were chosen. The first location was selected directly from GIS-generated random points. The second and third locations were determined using a random number generator to place them within ±1000 meters of the first point, ensuring sampling independence while staying within the fire perimeter and accessible areas. Points under 50 meters were regenerated to avoid spatial overlap.

### *2.4. Plant Survey*
Vegetation data were collected at each location along the trail using a 20 m transect. The fixed-radius point count was combined with the quadrat method to conduct a plant survey that accurately represented the overall vegetation composition of each site. At each location along the trail, a 20 m transect was established in a direction randomly determined by a number generator (0° to 360°). Three 35 cm by 35 cm quadrats were positioned along each transect at the 0 m, 10 m, and 20 m marks. For each quadrat, the direction (0° to 360°) and distance (0 m to 5 m) were randomly assigned using a number generator to ensure a fair representation of the site. Each quadrat was subdivided into a 3x3 grid, with each cell measuring 11.67 cm by 11.67 cm. The dominant plant species in each cell were identified based on their abundance and functional type. We conducted point counts within a 5 m radius of each quadrat to record the presence of large shrubs (height <1 meter and diameter <0.5 meters), trees, and herbaceous flowering plants. The functional categories included herbaceous flowering plants, grasses, and shrubs. According to the Royal Botanic Gardens Plants of the World Online and the University of California Berkeley Jepson Herbarium databases, herbaceous flowering plants were defined as non-woody species with flexible stems and an annual life cycle, while grasses were characterized as low, monocotyledonous plants belonging to the Poaceae family. Shrubs and trees, both woody plants, were differentiated by their branching patterns, shrubs with multiple basal branches and trees with a single erect trunk with branches higher up. Each categorization was validated through references in the aforementioned databases.

Environmental covariates, including elevation, slope direction, temperature, humidity, wind speed, air quality, soil texture, and time of day, were recorded. Weather data were sourced from The Weather Channel based on nearest station proximity; GPS and Apple Compass were used to obtain elevation and coordinates.

In cases of inaccessible or unsafe terrain (e.g., slopes >70°, dense vegetation, trail overlap), transect directions were rerandomized (~17% of transects). Similar rerandomization was applied to quadrat placement (~21% of cases) when locations fell on artificial surfaces or beyond feasible reach.

### *2.5 Bird Survey*
Bird surveys were conducted synchronously with plant surveys from 7 AM to 10 AM to coincide with peak avian activity. At each sampling point, a single observer conducted a 10-minute stationary point-count, recording all birds seen or heard within auditory range. The Merlin Bird ID application (Cornell Lab of Ornithology) assisted with sound-based species identification. Where species were ambiguous, observations were assigned to family or genus.

Birds were categorized as either specialists or generalists. Birds were considered specialists if more than one of the following criteria were met: (1) habitat range is exclusive and/or in fragmented distributions, (2) nesting and breeding require specific conditions, and/or (3) limited in diet options. If one or fewer criteria were met, then the bird species would be considered a generalist. Audubon.org and AllAboutBirds.org databases were used to identify habitat, nesting and breeding characteristics, and diets to determine their classification.

*2.6 Data Analysis*
To draw conclusions based on the collected data, analysis performed focused on the following focal points: (1) Changes in bird species richness or abundance across fire recovery stages in years, (2) Relationship between birds species and plant species in terms of both richness and abundance, (3) Impact of environmental covariates on bird species abundance and richness, and (4) Changes in bird and plant species abundance and diversity across fire bins.

To estimate whether bird species richness and abundance increased over time since fire, we used linear regressions and Pearson correlation analyses. To determine the impact of environmental covariates, such as weather, elevation, and temperature, on bird communities, we used linear regressions and ANOVA. The use of linear regressions allowed us to determine how a continuous predictor variable (time-since-fire) influences a continuous response variable (e.g., species richness and species abundance). ANOVA was used to test for statistical significance between different categories (time-since-fire).

Shannon index values were calculated for both bird and plant data to assess species diversity. The number of individuals within each unique species was tracked to calculate the Shannon index values. For the Shannon index values of plants, only point count data were included in the calculations. Quadrat cell data of plants were not included, as we only noted the dominant functional species and type present rather than individual counts.

All analyses were conducted in R (R Core Team 2023) and RStudio (Posit Team 2025). P-values were assessed at $\alpha = 0.05$. All code used for analysis is available in the supplemental materials.

3. Results

Across 14 chaparral sites, we recorded a total of 82 avian species and 1,437 individual birds. Species richness ranged from 7 to 29 per site, with the highest richness observed at a site last burned 22 years ago. Total abundance per site ranged from 33 to 156 individuals. Generalist species were present at all sites, while specialists were more frequently observed at sites with longer time since fire. The most commonly observed species included California Scrub-Jay (Aphelocoma californica), Mourning Dove (Zenaida macroura), and Spotted Towhee (Pipilo maculatus). Vegetation structure varied by site but was not quantified in this analysis.

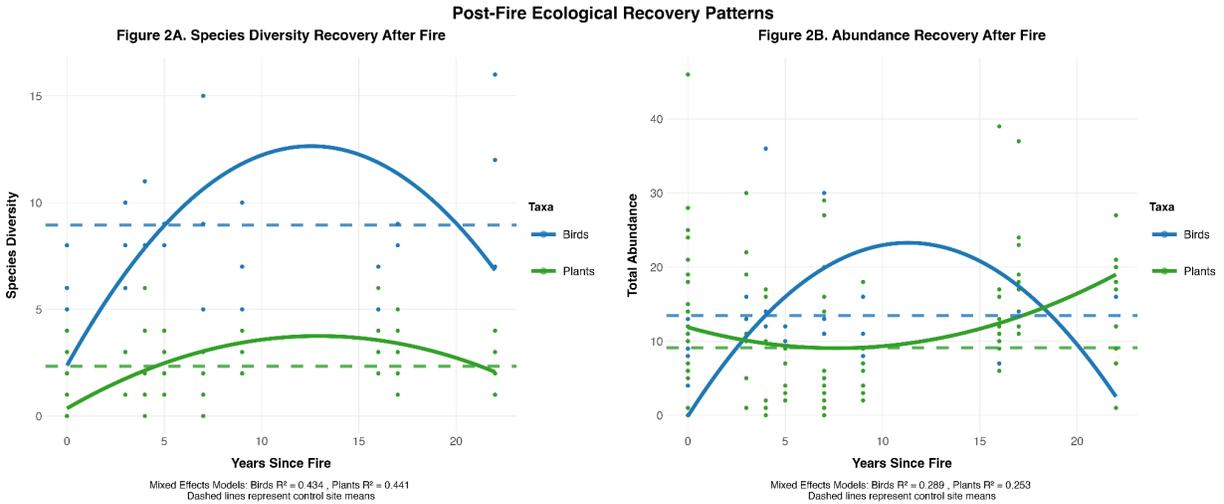

**Figure 2.** Time series graphs depicting bird and plant recovery patterns following wildfire disturbance. **Figure 2a** displays species diversity recovery trajectories, while **Figure 2b** shows total abundance recovery patterns over a 22-year post-fire chronosequence. Blue data points and fitted curves represent bird communities, while green represents plant communities. Solid curved lines represent mixed-effects quadratic regression models fitted to fire-affected site data (n = 945 observations). Horizontal dashed lines indicate mean values from unburned control sites (n = 162 observations) for reference. Mixed-effects models account for fire-specific random intercepts to control for variation among individual fire events (n = 10 fires). Model fit statistics are: bird diversity $R^2 = 0.434$, plant diversity $R^2 = 0.441$, bird abundance $R^2 = 0.289$, and plant abundance $R^2 = 0.253$.

To determine if plant and bird species richness and abundance could be correlated, a linear regression analysis was performed. We converted richness data to Shannon Diversity Index values to scale data between plant and bird species at the transect level. **Figure 3a** displays that plant richness was positively correlated with bird richness (r = 0.56, p = 0.0371), while **Figure 3b** visualizes that plant abundance showed no significant correlation with bird abundance (r = -0.043, p = 0.708).

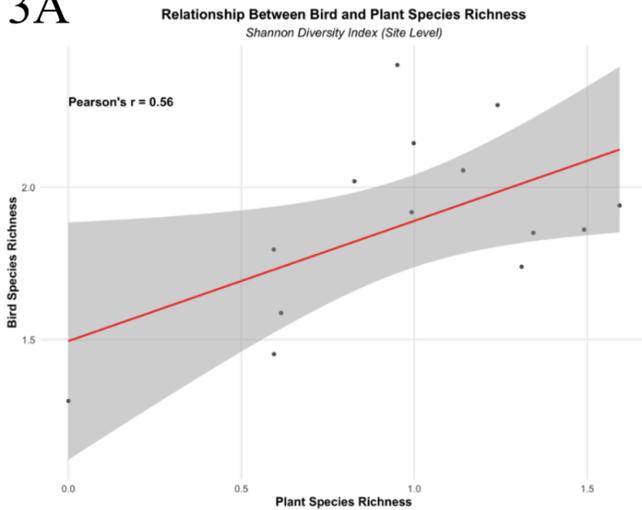
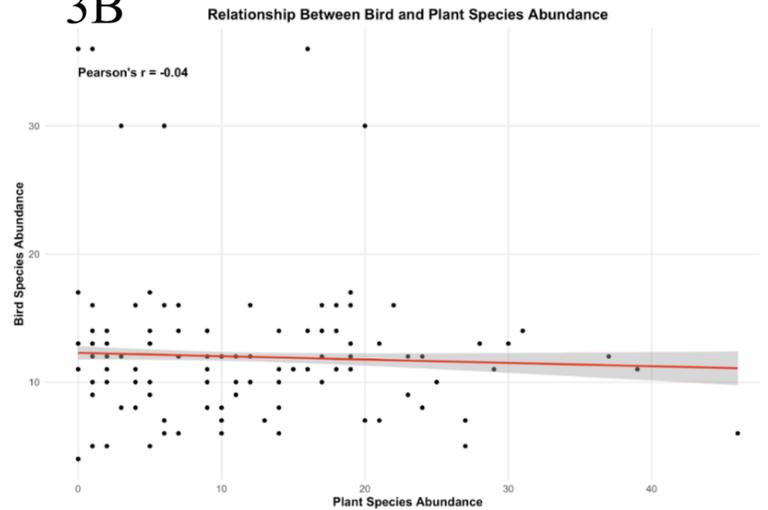

**Figure 3a.** *Relationship between bird and plant species richness across sites.* A positive correlation (r = 0.56, p = 0.0371) exists between plant and bird species richness based on Shannon Diversity Index values. Linear regression indicates that plant diversity explains ~31% of the variation in bird diversity ($R^2$ = 0.3141). **Figure 3b.** *Relationship between bird and plant species abundance across sites.* No statistically significant relationship was found between plant and bird abundance (r = -0.04, p = 0.1576). Only 0.18% of variation in bird abundance is explained by plant abundance ($R^2$ = 0.0018), suggesting other ecological drivers.

While a multitude of factors could be considered as covariates to our bird observations, we were able to account for Temperature, Elevation, and Weather. Additional factors in relative humidity and AQI were recorded, but some data were unable to be recorded and there was not enough variation in AQI to provide substantial results. **Figure 4** analyzes 3 abiotic factors on bird species abundance and richness to determine if there was an influence on our observations. Covariate analyses with a 95% confidence interval showed no correlation between weather (F = 0.361, p = 0.699), temperature (r = -0.181), or elevation (r = -0.251) on bird abundance. Similarly, there was no significant correlation between weather, temperature, or elevation on bird richness or diversity. Further, we aimed to characterize any differences in variation across fire bins, to determine if there were any changes to distribution as time since fire increased.

**Figure 4** ***Analysis of Covariates to Bird Abundance and Shannon Diversity Index:*** 4A-D used linear correlation analysis comparing Elevation in meters (A-B) and Temperature in Celsius (C-D). 4E and F compared variation across weather types using ANOVA analysis, where weather groups were either Cloudy, Foggy, or Sunny. Red is the correlation line where the correlation

coefficient was calculated. Elevation: r = −0.251 (95% CI = −0.497 - 0.0319) and r = −0.224 (95% CI = −0.475 - 0.0612) for abundance and diversity, respectively. Temperature: r = −0.181 (95% CI = −0.440 - 0.105) and r =−0.00897 (95% CI = −0.273 - 0.289). Weather: F = 0.361 (p>0.05) and F = 0.00259 (p>0.05).

We wanted to characterize the changes in plant composition within the sites that we were surveying. Many bird species, especially taxa that prefer particular habitats, may venture into locations with ideal functional types to their behavior. The functional types that colonize areas post-fire change over time, and analyzing how the composition of vegetation changed over fire recovery stages could help explain the connection between bird richness and abundance changes. The average abundance and species richness of the 3 plant functional groups were compared at each fire bin level.

**Figure 5** reveals distinct successional strategies among plant types following fire disturbance. Herbaceous plants exhibit classic pioneer behavior, dominating abundance immediately post-fire (8.21 individuals, 0.71 species) before declining precipitously as woody vegetation establishes. Shrubs demonstrate the most consistent recovery trajectory, with both diversity and abundance increasing steadily from 0.42 species and 2.21 individuals at 0-2 years to 2.56 species and 14.33 individuals by 20+ years, eventually comprising the dominant component of recovering plant communities. Trees show delayed establishment, remaining minimal through the first decade (≤0.72 species) but peaking in abundance during the 10-20 year period (3.39 individuals) before declining again, suggesting competitive exclusion by shrubs in later succession. The temporal mismatch between peak total diversity (3.11 species at 10-20 years) and peak total abundance (17.63 individuals at 20+ years) indicates that community assembly and biomass accumulation follow different trajectories. This pattern supports the intermediate disturbance hypothesis for diversity while demonstrating that abundance continues to increase as shrub-dominated communities mature, ultimately creating plant communities structurally dominated by shrubs rather than the mixed herb-shrub-tree assemblages that characterize peak diversity periods.

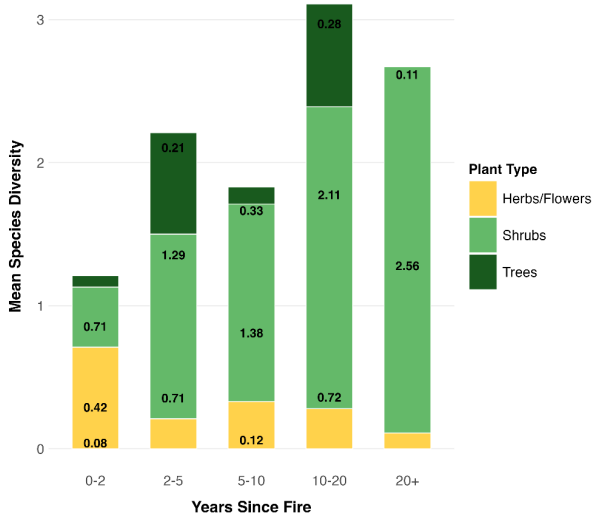
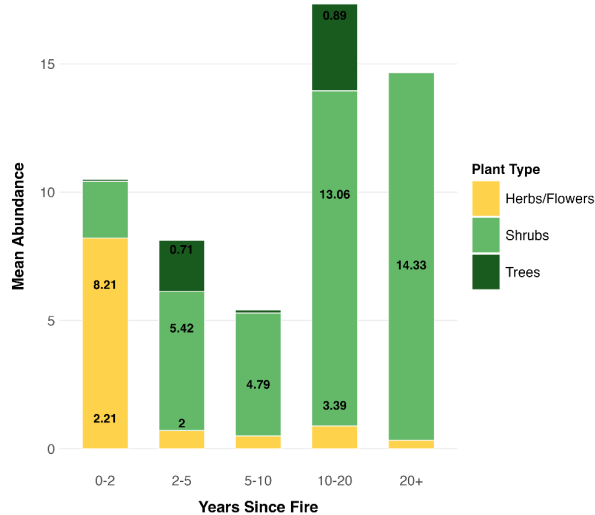

**Figure 5.** Temporal patterns in plant functional types across fire recovery stages. Stacked bar charts display mean species diversity (**Figure 5a**) and mean abundance (**Figure 5b**) for trees (dark green), shrubs (light green), and herbaceous/flowering plants (yellow) across five post-fire recovery periods (n = 216 observations for 0-2, 2-5, and 5-10 year periods; n = 162 for 10-20 years; n = 81 for 20+ years). Values represent means calculated from fire-affected sites, with exact values displayed on each bar segment. Trees and shrubs increased steadily with time, while herbaceous species peaked immediately post-fire.

**Figure 6** uses a similar bar graph to show how generalist bird species and specialists changed in abundance and richness in the same periods as average plant functional types. Average generalist bird counts remained relatively constant and dominant in the earlier stages, though the later fire stages appear to have increased variation (5-10 years; SE: 0.889) (10-25 years; SE: 0.873) compared to earlier fire stages (0-2 years; SE: 0.512) (2-5 years; SE: 0.378) and control sites (SE: 0.577). Specialists appear to increase as fire stages progress, where the later and unburned stages show the highest average abundances (10-25: 3.67, Control: 3.83) and the most recent fires had the minimum average abundance (1.44).

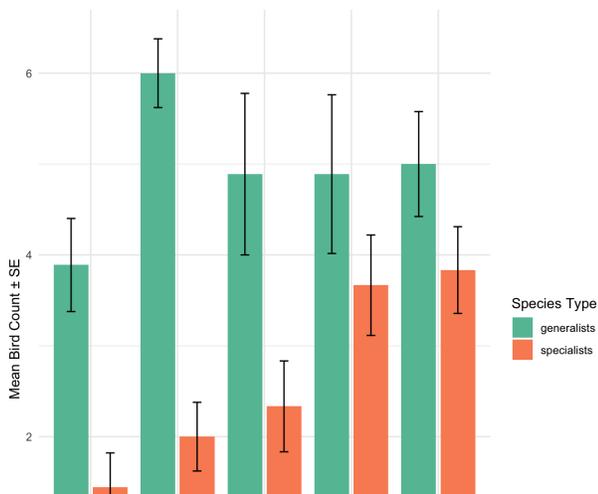

**Figure 6.** *Proportion of generalist vs. specialist bird species across fire recovery stages.* Mean bird count abundances, accompanied by standard error bars, at fire bin level visualized by generalist species in orange and specialist species in green. Both generalist and specialist species

demonstrated a decrease in average abundance directly following burning, with generalist species peaking 2-5 years after fire and specialist species showing gradual recovery and returning to near unburned (control) abundance levels at 10-25 years after fire.

## 4. Discussion

Our study set out to explore how bird communities respond to fire-driven changes in chaparral habitats over time. The 14 study sites across Los Angeles County revealed significant differences in bird richness and abundance across the post-fire chronosequence, with more recently burned sites consistently supporting lower diversity. As vegetation matured, we observed an increase in both species richness and ecological specialization, consistent with successional theory and prior research in Mediterranean-climate ecosystems (Clavero et al. 2011; Hargrove and Unitt 2018). Bird richness in recently burned areas was noticeably low, with groupings dominated by generalist species capable of exploiting simplified environments (Mendelsohn et al. 2008, Clavero et al. 2011, Hargrove and Unitt 2018). These early successional sites were characterized by sparse shrub cover and reduced vertical stratification, structural features known to limit the availability of microhabitats for nesting and foraging. Our results found similar patterns to those described by previous researchers; for example, MacArthur and MacArthur (1961) demonstrated that bird species richness increases with vegetation structural complexity, particularly benefiting habitat specialists, a finding consistent with our observation that older, structurally mature sites supported more specialized bird communities (Moreira et al. 2003).

Interestingly, richness peaked not at the oldest sites but in some intermediate-aged locations, suggesting a nonlinear recovery trajectory. This is consistent with the intermediate disturbance hypothesis (Connell and Slatyer 1977), which posits that species diversity is often maximized at intermediate levels of disturbance due to the coexistence of pioneer and late-successional species. Our observations of elevated richness and structural heterogeneity in mid-successional habitats align with this framework. These mid-successional habitats exhibited a mix of open and closed vegetation patches, potentially supporting a wider array of ecological niches. This transitional phase appears to represent an important window for species coexistence, where generalists persist alongside returning specialists. Such habitat heterogeneity has been shown to support higher biodiversity across taxonomic groups, reinforcing the value of landscape mosaics for conservation (Tews et al. 2004, Stein et al. 2014).

However, raw richness values alone did not always reflect ecosystem quality. In some cases, elevated richness was driven by a proliferation of generalist species with broad ecological tolerances, while older sites, although sometimes lower in richness, supported communities with a higher proportion of obligate chaparral specialists. This illustrates a central critique in biodiversity monitoring: richness, while more intuitive than abundance and more easily measured, may tell an incomplete story about degradation in community composition (Julliard et al. 2006, Gregory and van Strien 2010). Assessments that account for functional traits and species turnover are necessary to reveal whether communities are truly recovering or simply reassembling in a less specialized state.

Our results also emphasized the ecological consequences of invasive grasses, especially Bromus species. These grasses were prevalent at many of our study sites and likely contributed to delayed or incomplete shrub regeneration (Park and Jenerette 2019, Chambers et al. 2019). By outcompeting native vegetation, Bromus fosters habitat simplification and may perpetuate a fire-prone grass–fire cycle that prevents chaparral from reestablishing (Park and Jenerette 2019, Palit and DeKeyser 2022). In our survey, sites with substantial Bromus cover were consistently associated with reduced vegetation complexity and lower richness of specialist birds, corroborating previous research on type conversion and community collapse in fire-adapted systems (Pratt 2022).

Spatial variability in recovery among sites of similar fire age has been attributed in the literature to abiotic conditions such as slope, aspect, soil depth, and post-fire precipitation (Mendelsohn et al. 2008, Seavy and Alexander 2014). However, we did not detect consistent effects of these factors in our data. This likely stems from our sampling design, which involved only a single visit to each site. As a result, we were unable to observe how temporal variation in conditions such as wind, temperature, or cloud cover may have influenced bird activity or detectability. This limitation restricted our ability to assess the impact of fine-scale abiotic variability on bird communities and should be addressed in future studies with repeated site visits and broader temporal coverage. Although previous studies have found that north-facing slopes with greater moisture retention support faster shrub development and more diverse bird communities, while south-facing slopes remain open and grass-dominated for longer periods, we did not observe such patterns in our data (Zahura et al. 2024). Nevertheless, the literature suggests that managing for heterogeneity, rather than uniform restoration timelines, can yield more ecologically robust outcomes (Tingley et al. 2016). This contrasts with earlier studies that emphasized uniform recovery rates across post-fire landscapes (e.g., Pausas and Keeley 2009), suggesting that local abiotic variation may play a more dominant role than previously assumed in structuring post-fire avian communities, rather than uniform restoration timelines, will yield more ecologically robust outcomes (Branco et al. 2015).

Another important consideration is species turnover across the chronose sequence. Some generalist species present in early successional sites disappeared from older stands, potentially suggesting a successional filtering process where species with low habitat specificity are replaced by specialists better adapted to complex environments (Tingley et al. 2016). However, a few generalists persisted throughout the gradient, indicating that certain flexible foragers can exploit a range of structural contexts (Sweeney et al. 2010). The persistence of these species may mask underlying declines in habitat specialists and alter the functional balance of communities over time. This phenomenon parallels findings by Devictor et al. (2008), who documented biotic homogenization across bird communities in disturbed European landscapes. In such cases, although species richness may appear stable, the replacement of habitat specialists by generalists leads to communities that are functionally degraded and increasingly similar across different ecosystems. This hidden simplification reduces ecological uniqueness and may compromise ecosystem resilience and function.

Our study also highlights the critical role of vertical stratification in supporting diverse avian assemblages. Vegetation layers (ground cover, mid-story shrubs, and canopy elements) create microhabitats that fulfill different life-history requirements, from foraging to nesting and predator avoidance (MacArthur and MacArthur 1961, Martin and Finch 1995). Sites with more defined vertical layering consistently supported a greater number of species and higher evenness in relative abundance. Sites with more defined vertical layering consistently supported a greater number of species and higher evenness in relative abundance, a pattern also observed by MacArthur and MacArthur (1961) and reinforced by Díaz et al. (1998), who linked habitat complexity to avian diversification across Mediterranean woodlands. Recreating the structural complexity necessary to support a diverse avian community is not easily done without involved conservation, emphasized the need for long-term planning and targeted planting or assisted regeneration (e.g., active replanting of native shrubs or facilitating natural recruitment through targeted removal of competitive invasives or ground preparation) to promote an increase in vertical stratification in severely degraded sites (Crouzeilles et al. 2020).

In terms of fire management implications, our findings support maintaining fire return intervals that allow full chaparral recovery (Davis and Michaelsen 1995). While fire suppression remains essential for protecting life and property, it should not result in excessive fuel buildup that increases the risk of high-intensity megafires. Conversely, overly frequent burns (often linked to human ignitions; Nagy et al. 2018) can interrupt successional processes and degrade native seed banks, facilitating the spread of invasive species. Effective land management must balance the need to reduce fire risk with the preservation of natural regeneration cycles. Based on our observations, sites that had not burned for at least 15–20 years consistently supported more diverse and compositionally mature bird communities. This suggests that late-successional avian assemblages may require extended intervals to fully reestablish, likely reflecting the time needed for structural vegetation features, such as dense shrubs and vertical layering, to develop. However, this threshold likely varies depending on local ecological conditions. To improve ecological outcomes, post-fire restoration efforts should incorporate invasive species control, particularly targeting Bromus during early regeneration phases (Devendra Dahal et al. 2021, Fusco et al. 2022). Mechanical removal, grazing, and controlled burns have shown some success in reducing grass dominance, but these interventions must be carefully timed and site-specific to avoid collateral damage (Davies et al. 2016, Chambers et al. 2024). Restoration should also consider planting native shrubs or seeding with native forbs to accelerate recovery and resist invasion (Halassy et al. 2023, Arychuk et al. 2024). Monitoring protocols should integrate avian community metrics as indicators of structural habitat success (Adler and Jedicke 2022). Future research should expand on the chronosequence design by incorporating more precise vegetation metrics, such as remote-sensed LiDAR data or canopy photogrammetry, to quantify structure at multiple scales (Jacon et al. 2024). Studies could also benefit from a behavior or trait-based approach to bird community analysis, exploring how specific foraging strategies or nesting preferences correlate with habitat variables, potentially even using an ethogram. Additionally, coupling bird surveys with data on reproductive success would provide stronger inference on habitat quality.

Another promising direction lies in studying community resilience following repeat burns. As fire frequencies increase, many chaparral stands are experiencing their second or third burn within a few decades (Grupenhoff and Safford 2024). Syphard et al. (2019) found that fire frequencies in Southern California chaparral have increased significantly due to human ignitions, with many stands burning multiple times within a few decades. Their findings show that short fire intervals (particularly those under 15 years) are strongly associated with the loss of woody cover and widespread vegetation type conversion to herbaceous dominance. This supports our own observations of reduced structural complexity and specialist bird richness in more frequently burned areas, emphasizing the threat repeated fires pose to both plant and bird community recovery. Understanding how these cumulative disturbances affect bird communities, whether through compounding habitat loss, reduced seed banks, or altered predator-prey dynamics, will be essential for anticipating long-term biodiversity trends.

Lastly, while our study focused on birds, similar successional dynamics likely apply to other taxa, including mammals, reptiles, and invertebrates. Cross-taxa comparisons could reveal broader ecosystem trends and help identify conservation strategies that benefit multiple groups simultaneously. Multispecies monitoring programs, especially those linked to vegetation recovery indices, could be instrumental in this effort.

Our findings highlight the intricate relationship between vegetation structure and bird community dynamics in post-fire chaparral. While richness tends to recover over time, true ecological recovery hinges on the return of specialists and the re-establishment of complex vegetative structures. Through targeted management and long-term ecological monitoring, chaparral systems can be guided toward resilient and functionally diverse futures in the face of growing fire risk and ecological uncertainty.

## 5. Acknowledgments

We would like to express our gratitude and acknowledgement to Dr. Elsa Ordway and Dr. Felipe Zapata for giving us invaluable advice and guidance throughout our field quarter. We thank Auxenia Grace Privett-Mendoza and Lindsay Reedy for their feedback and unwavering support of our research. We would like to show our appreciation to Kristen Ochoa for providing us with her insight and access to the Rubio Canyon in Altadena. Lastly, we'd like to acknowledge that our sites across Los Angeles County, are located on the traditional, ancestral, and unceded territory of the Gabrielino-Tongva, Chumash, Fernandeño Tataviam, Ventureño Chumash, and Yuhaaviatam/Maarenga'yam (Serrano) Tribes.